\documentclass[conference]{IEEEtran}
\IEEEoverridecommandlockouts
\usepackage{algorithm}
\usepackage{algorithmic}
\usepackage[mode=buildnew]{standalone}
\usepackage{amsmath,mathtools,nccmath}
\usepackage{siunitx}
\usepackage{cite}

\usepackage{dblfloatfix}
\usepackage{booktabs}
\usepackage{multirow}

\usepackage{enumitem}
\usepackage{mathtools}
\usepackage[utf8]{inputenc}
\usepackage{amssymb}
\usepackage{gensymb}

\usepackage[compact]{titlesec}
    \titlespacing{\section}{0pt}{2ex}{1ex}
    \titlespacing{\subsection}{0pt}{1ex}{0ex}
    \titlespacing{\subsubsection}{0pt}{0.5ex}{0ex}
    \usepackage[]{footmisc}
\usepackage{float}
\usepackage{array}
\usepackage{algorithmic}
\usepackage{textcomp}
\usepackage{xcolor}
\usepackage{tabularx}
\def\BibTeX{{\rm B\kern-.05em{\sc i\kern-.025em b}\kern-.08em
    T\kern-.1667em\lower.7ex\hbox{E}\kern-.125emX}}

\begin{document}

\include{bibliography}

\title{A Hybrid Three-Port Topology for Urban Charging Stations\\
}

\author{
\IEEEauthorblockN{Mohammadreza Khodaparast Klidbari$^1$,  Naser Souri$^2$, Zahra Sadat Habibolahi$^3$, Hamid Montazeri Hedeshi$^4$}

\IEEEauthorblockA{ $^1$Faculty of Electrical Engineering, K. N. Toosi University of Technology, Tehran, Iran.  \\ $^{2}$The Bradley Department of Electrical and Computer Engineering, 
Virginia Tech, United States \\ $^3$Department of Electrical Engineering, Iran University of Science and Technology, Tehran, Iran \\ $^{4}$Faculty of Electrical and Computer Engineering, University of Tabriz, Tabriz, Iran\\
%\IEEEauthorblockA{$^2$Department of Energy and Control, University of Normandy-ESIGELEC-IRSEEM, Rouen, France}
Emails: $^1$khpmohammadreza@email.kntu.ac.ir, $^2$nsouri@vt.edu, $^3$zahra\_habibolahi78@elec.iust.ac.ir, $^4$h.montazeri.h@gmail.com} %, tehrani@esigelec.fr}
 \thanks{“© 20XX IEEE.  Personal use of this material is permitted.  Permission from IEEE must be obtained for all other uses, in any current or future media, including reprinting/republishing this material for advertising or promotional purposes, creating new collective works, for resale or redistribution to servers or lists, or reuse of any copyrighted component of this work in other works.”}
}

\maketitle

\begin{abstract}

Electric vehicles are rapidly gaining popularity as a sustainable alternative to conventional gasoline. In urban areas, chargers with different ratings can accommodate the diverse needs of electric vehicles. However, the available multiport topologies have variable switching frequencies. This paper introduces a hybrid multiport isolated DC-DC converter for urban charging stations, incorporating fast and slow charging ports with a fixed switching frequency. It provides isolation and enables soft switching on the primary side of the converter without circulating current on its secondary side. The primary side does not need feedback, which reduces complexity. The second stage generates a wide output voltage range to charge the electric vehicle battery by employing a switch. In addition, the proposed topology offers reduced component count and simple control with fixed-frequency operation. This paper provides the concept and the operation modes. Experimental results are provided to validate its features. The prototype converter achieves 96\% peak efficiency.

\end{abstract}


%=======================================================
\section{Introduction}
\begin{figure}[!t]
\centerline{\includegraphics[width= 0.47\columnwidth ]{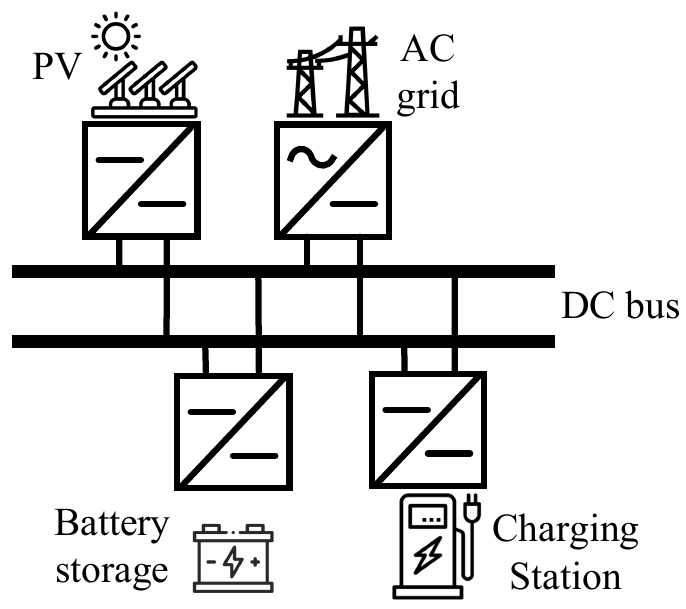}}

\vspace*{-0.3 cm}
\caption{Charging station topology.}
\label{fig:station}

\vspace*{-0.4 cm}
\end{figure}

Increasing greenhouse gas emissions from the automotive industry is a concerning issue. Electric and hybrid electric vehicles have received significant attention in recent years with advancements in battery technology \cite{Chevinly, somaz}. Their potential to lower greenhouse gas emissions and lessen reliance on fossil fuels promotes sustainable transportation. Therefore, charging stations are increasing in number in urban areas. Charging stations can benefit from an AC or DC grid. Fig.~\ref{fig:station} shows a DC-based charging station structure supplying from both AC and DC sources. DC grids have fewer power conversion stages than AC grids, reducing power conversion losses and potentially increasing efficiency as electric vehicle batteries are also powered with DC \cite{naser, Mahdi, 9767460}.
%In addition, the DC structure offers less complicated control and ease of design. 

\begin{figure}[!t]
\centerline{\includegraphics[width= 0.97\columnwidth ]{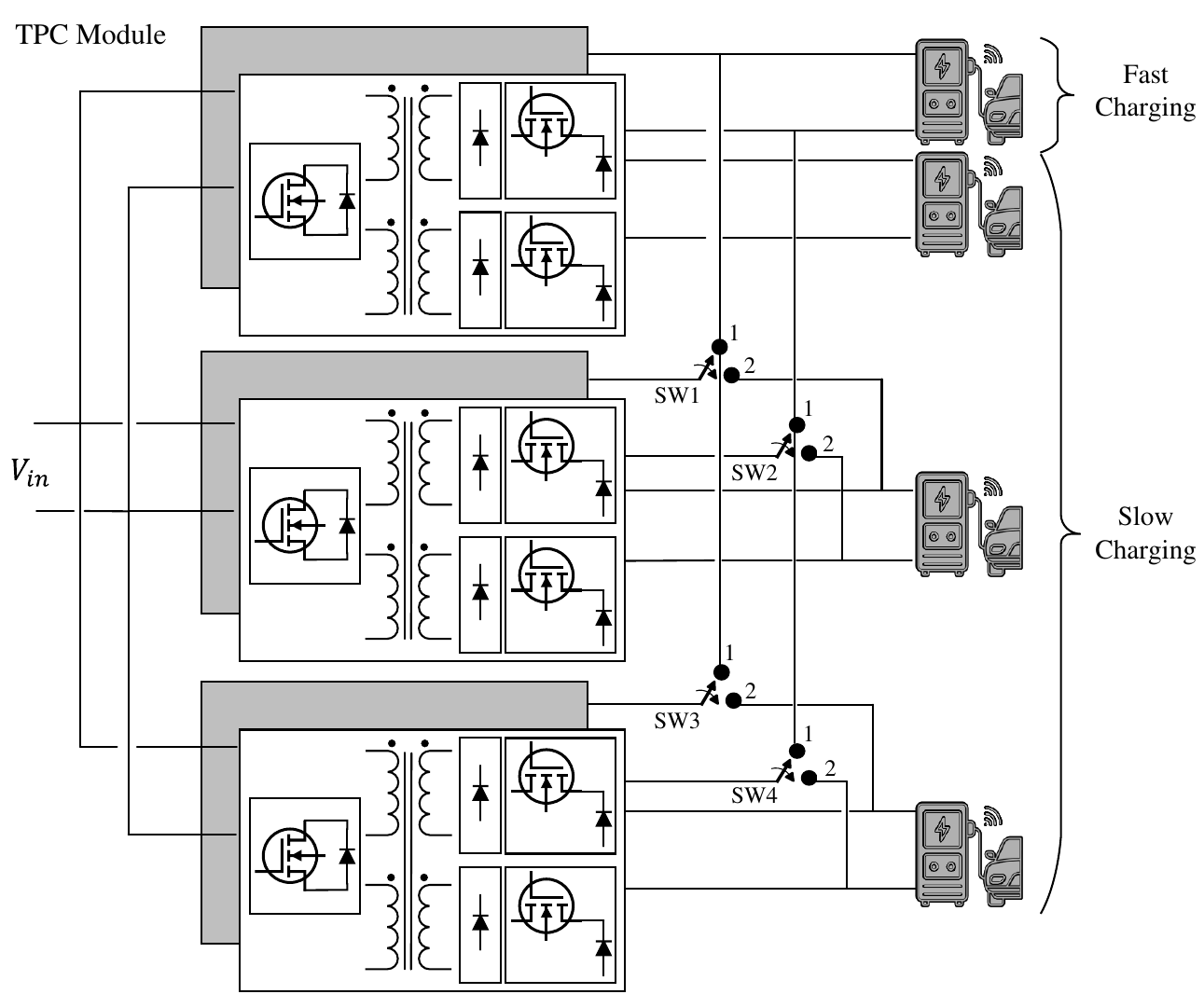}}

\vspace*{-0.4 cm}
\caption{A hybrid charging station block diagram.}

\vspace*{-0.4 cm}
\label{fig:hybrid}
\end{figure}

Slow charging stations, commonly known as level 2, are ideal for overnight charging or suitable for parking periods, while fast charging is ideal for long journeys and rapid energy replenishment in suburban areas, ensuring that drivers can conveniently and efficiently charge their vehicles. Therefore, hybrid charging stations enhance flexibility, accommodating various charging speeds for a diverse range of EVs \cite{Assadi, Krishnaswami}. Hybrid charging stations can benefit from multiport topologies, as they can integrate fast and slow chargers in one package, as shown in Fig. \ref{fig:hybrid}. In addition, multiport topologies usually offer a lower number of components, which reduces the cost. 
%Therefore, hybrid charging stations have emerged as a good solution to accommodate the diverse needs of electric vehicles.
%Slow charging stations, commonly known as Level 2 chargers, are ideal for overnight charging and suitable for parking periods. 

%Hybrid charging stations could benefit from multiports with multi rate topologies. These types of topologies enable the integration of fast and slow chargers, which increases the flexibility of charging stations by accommodating varying charging speeds for a diverse range of EVs. Multiport topologies offer lower number of components, which reduce the cost. 
Among multiport topologies, LLC provides a simple structure with zero-voltage switching (ZVS) capability on its primary side and a wide range of output voltage\cite{Kougioulis}. ZVS ensures that the voltage across the switch is zero when it turns on or off, reducing the switching losses. The main challenge for LLC topology is that the switching frequency is not fixed and varies according to the output voltage. On the other hand, to achieve a high gain, the switching frequency should be less than the resonant frequency. However, reducing the switching frequency affects the transformer and inductor size. In contrast, considering a higher switching frequency increases the switching losses. In addition, adjusting the voltage gain using the switching frequency complicates the transformer design and can lead to a reduction in efficiency \cite{Ngoc, Mohebifar}. 
% \cite{8930619} 

For EV applications, it is recommended that the input and output are galvanically isolated so that battery protection is not affected by the charging system. However, \cite{7590135,8558208} propose topologies that isolation is not respected. References \cite{Krishnaswami, Wang,Phattanasak} present an isolated three-port topology with two LLC resonant tanks that can operate at a high switching frequency with lower switching losses due to soft-switching operation. However, the design of these topologies is complex as they use a three-winding transformer. In addition, this type of transformer introduces a circulating current, which increases the losses and stress over the components. Therefore, an additional control algorithm is required to minimize this current. Reference \cite{Shi } proposes a topology that the frequency is fixed during some modes. However, it changes in other modes. Therefore, \cite{Fuxin} proposes a new topology that uses a fixed frequency. However, this topology has a single port structure. Moreover, it has a large number of components, which increases the cost. To overcome the existing drawbacks, a new multiport SiC-based isolated topology is proposed in this paper to address the aforementioned issues. Table I shows a comparison of different topologies. This comparison is based on the control method, the number of ports, the circulating current, the number of components, and the bidirectional or unidirectional structure. 
%For example, the proposed topology is multiport and there is no circulating current in the circuit. However, the proposed topology in [90] is single-port and the circulating current is high, which cause stress on the components and additional losses. 

\begin{table*}[t]
\caption{Comparison of Different Topologies.}

\vspace*{-0.5 cm}
\begin{center}
\begin{tabular}{llllllllcc}
\toprule
{Reference} & Control method & Circulating current & Port & Switching & Direction & Transformers & Inductors  & Diodes  &  Switches  \\
\midrule

Proposed & Fixed $F_\text{sw}$, PWM & No  & Multiport & ZVS  & Unidirectional  & 2  & 4  &  8  &  6  \\
\midrule

\cite{Fuxin} & Variable $F_\text{sw}$, PWM & No & Single port & ZVS  & Unidirectional  & 3  & 3  &  8  &  6  \\
\midrule

\cite{Gao} & Variable $F_\text{sw}$, PSH & Yes  & Multiport & ZVS  & Unidirectional  & 2  & 2  &  4  &  6  \\
\midrule

\cite{Yong} & Variable $F_\text{sw}$, PWM & Yes  & Multiport & ZVS  & Unidirectional  & 1  & 2  &  8  &  8  \\
\midrule

\cite{9479082} & Fixed $F_\text{sw}$, PS-PWM & Yes  & Multiport & Hard  & Unidirectional  & 1  & 1  &  8  & 8   \\
\midrule

\cite{Chakraborty} & Variable $F_\text{sw}$, PSH & No  & Multiport  & ZVS  & Unidirectional  & 1  & 2  &  0  &  12  \\

\bottomrule
\end{tabular}
\label{table:system_parameters}
\end{center}

\vspace*{-0.2 cm}
\end{table*}

The contribution of this paper is as follows.
\begin{itemize}
  \item Proposing a new single-stage multiport topology for hybrid charging stations with fixed frequency.
\end{itemize} 

Section II provides the proposed topology, its features, and its operation modes. Section III provides simulation studies to evaluate the switching waveforms. Section IV provides experimental studies to validate the effectiveness of the proposed topology. In the last section, a conclusion is provided.

%=======================================================
\section{Proposed Multiport Topology}
\begin{figure}[t]
\centerline{\includegraphics[width= 0.99\columnwidth ]{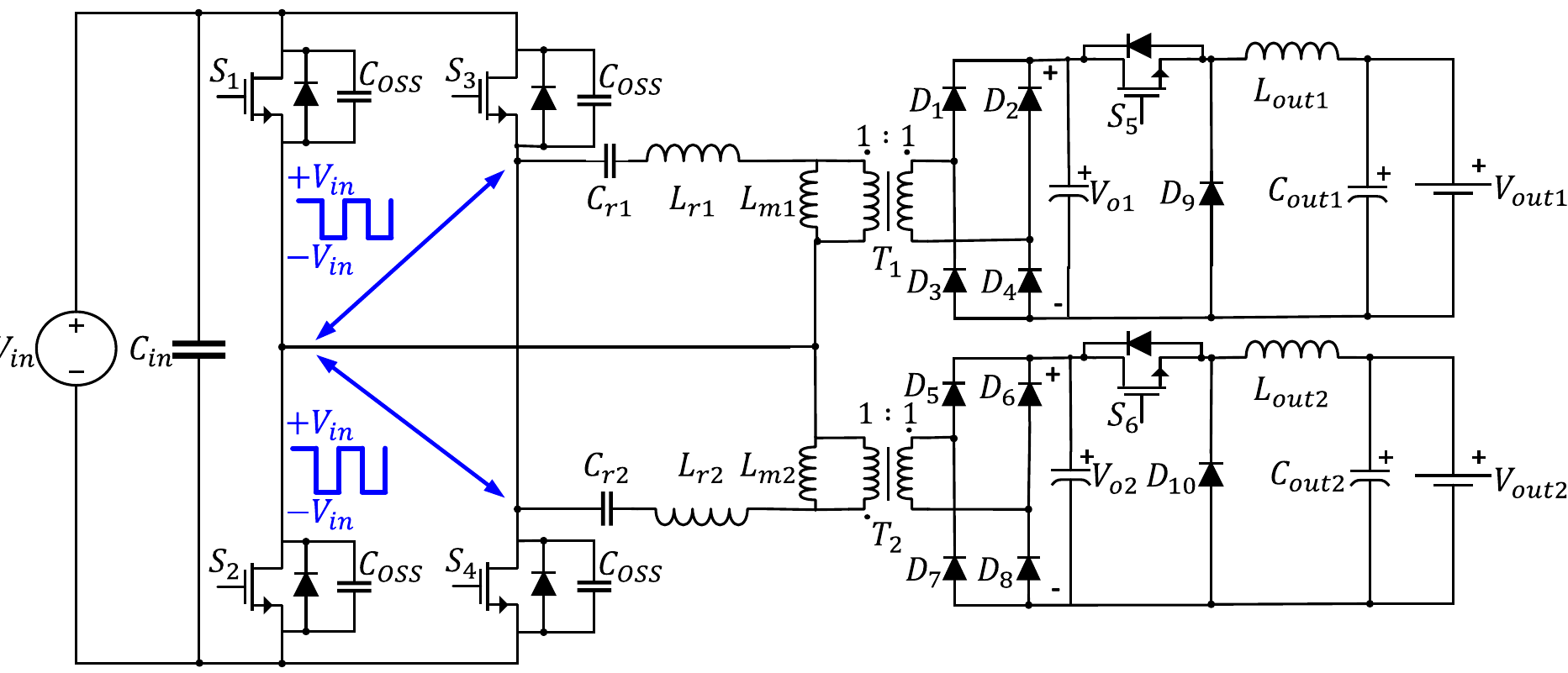}}

\vspace*{-0.4 cm}
\caption{Proposed multiport LLC topology for a hybrid charging station.}
\label{fig:topology}

\vspace*{-0.3 cm}
\end{figure}

This section studies the concept of the proposed charger topology and the operating modes. Fig.~\ref{fig:topology} shows the proposed LLC topology for a charging station. This topology consists of two stages and two outputs, one of which is fast charging, and the other is slow charging. The primary side of this converter is a full bridge converter, and the switching frequency is close to the resonance frequency. The transformer isolates the primary and secondary sides. The power is shared between the secondary sides through this transformer. ZVS occurs for the primary switches, reducing the switching losses. However, the charging algorithm is implemented on the secondary side. The proposed topology utilizes a two-winding transformer instead of a three-winding transformer, resulting in eliminating the circulating current and reducing complexity. To achieve ZVS during dead time, the magnetizing current of the transformer $I_m$ must be capable of charging and discharging the capacitors $C_\text{oss}$. The current that passes through each of the output capacitors $C_\text{oss}$ is calculated as follows.

\begin{equation}
I_\text{coss}= C_\text{oss}\frac{\Delta V}{\Delta t}
\end{equation}

\noindent where $I_\text{coss}={(I}_\text{lm1}+I_\text{lm2})/2$ and $I_\text{lm1,2}=V_\text{out1,2}/4L_mf_s$, and $f_s$ is the switching frequency. Therefore, the duration for charging the capacitor $C_\text{oss}$ should be less than the dead time to achieve zero-voltage switching. Fig.\ref{fig:operation} shows the circuit in different operation modes for the proposed topology. These modes are discussed below.

%Fig.6 illustrates the equivalent circuit during dead time when the converter operates close to the resonant frequency. To achieve ZVS during dead time, the magnetizing current of the transformer $I_m$ must be capable of charging and discharge of capacitors $C_\text{oss}$. When using two transformers, The magnetizing currents of the transformers charge and discharge the capacitors. The following equation calculate the current that passes through each of $C_\text{oss}$

\subsection{Operation Mode I: $t_1\leq t < t_2 $}
Fig. \ref{fig:operation}(a) shows the equivalent circuit in this operation mode during $t_1$ to $t_2$. In this mode, $S_2$ and $S_3$ are ON at the primary side. At the beginning of this interval, the voltage across the MOSFET is zero as the body diode conducts before $t_1$, which indicates a ZVS. In this mode, $I_\text{lr1}$ and $I_\text{lr2}$ are positive and greater than $I_\text{lm1}$ and $I_\text{lm2}$. As a result, diodes $D_1$,$D_4$,$D_5$,$D_8$ are ON and $V_\text{lm1}$ and $V_\text{lm2}$ are equal to $V_{o1}$ and $V_{o2}$, respectively.

\subsection{Operation Mode II: $t_2\leq t < t_3 $}
In this mode, $S_2$ and $S_3$ are ON and $S_1$ and $S_4$ are OFF as shown in Fig. \ref{fig:operation}(b). In this mode, $I_\text{lr2}$ reaches $I_\text{lm2}$ at $t_2$, and the secondary side of $T_2$ is zero. In this case, the magnetizing inductor at the primary side of this transformer resonates with the resonant tank, which is calculated from $f_\text{sr2}=\frac{1}{2\pi\sqrt{(l_r+l_m)c_r}}$.

\subsection{Operation Mode III: $t_3\leq t < t_4 $}
In this mode, all switches are OFF, as shown in Fig. \ref{fig:operation}(c). $I_\text{lr1}$ is greater than $I_\text{lm1}$ in this mode, helping $I_\text{lm1}$ to charge and discharge the capacitors $C_\text{oss}$ to achieve ZVS. The resonant tank current passes through the output capacitor of the switches, and based on the current direction, it charges the capacitor of the switches $S_2$ and $S_3$, and discharges $S_1$ and $S_4$. Finally, the capacitors $C_\text{oss}$ are fully charged and discharged, and the current passes through the body diodes $S_{1,4}$.

\subsection{Operation Mode IV: $t_4\leq t < t_5 $}
In this mode, the body diodes of $S_1$ and $S_4$ are ON, and $D_1$ and $D_2$ conduct as shown in Fig. \ref{fig:operation}(d). The direction of the inductors current remains unchanged during this interval. In addition, $I_\text{lr1}$ and $I_\text{lr2}$ pass through the body diodes of $S_1$ and $S_4$. The $V_\text{DS}$ is, therefore, zero for $S_1$ and $S_4$.

\subsection{Operation Mode V: $t_5\leq t < t_6 $}
\begin{figure}[!t]
\centerline{\includegraphics[width= 0.98 \columnwidth ]{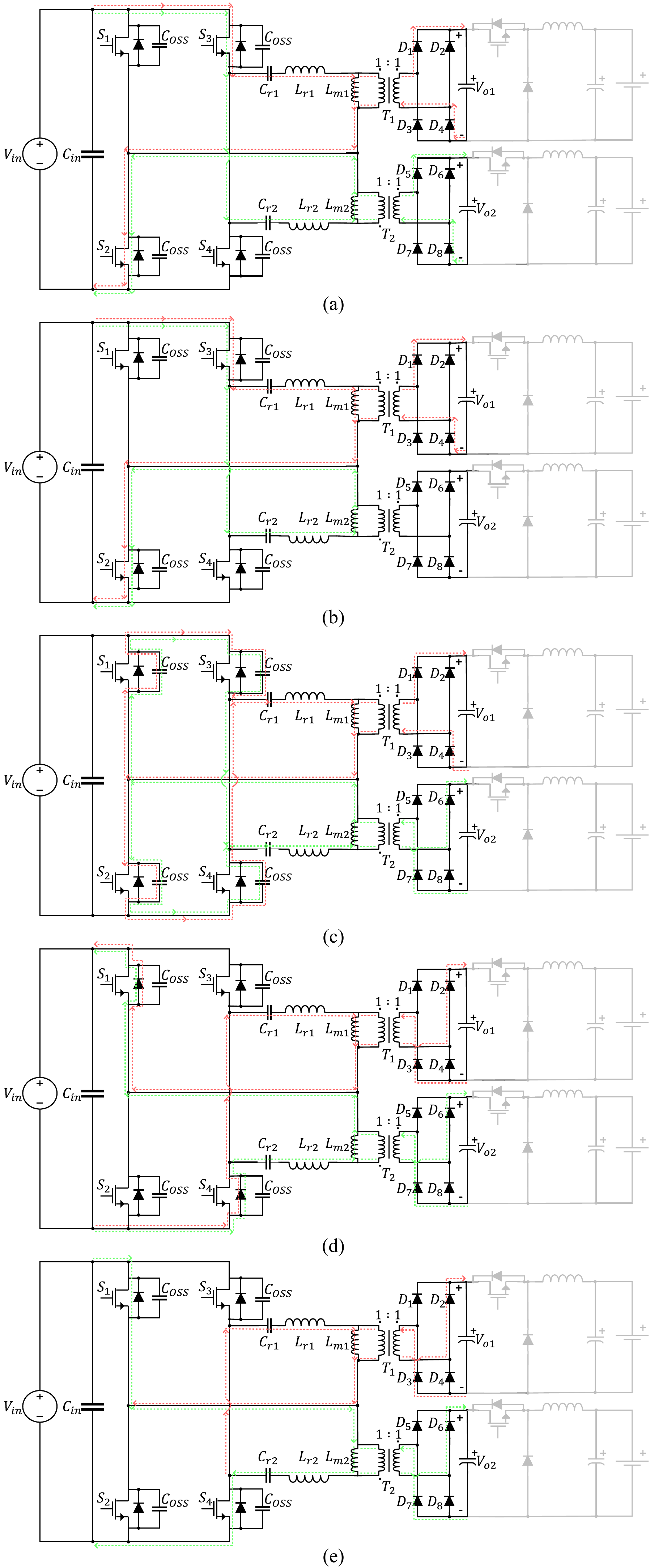}}

\vspace*{-0.3 cm}
\caption{Operation modes of the topology: (a) mode I, (b) mode II, (c) mode III, (d) mode IV, (e) mode V.}

\vspace*{-0.5 cm}
\label{fig:operation}
\end{figure}

In this mode, $S_1$ and $S_4$ are ON and $D_2$, $D_3$, $D_6$, and $D_7$ conduct as shown in Fig. \ref{fig:operation}(e). In the previous mode (IV), the body diodes of $S_1$ and $S_4$ are ON. Therefore, the corresponding switches turn on with ZVS.

\begin{figure}[t]
\centerline{\includegraphics[width= 0.98 \columnwidth ]{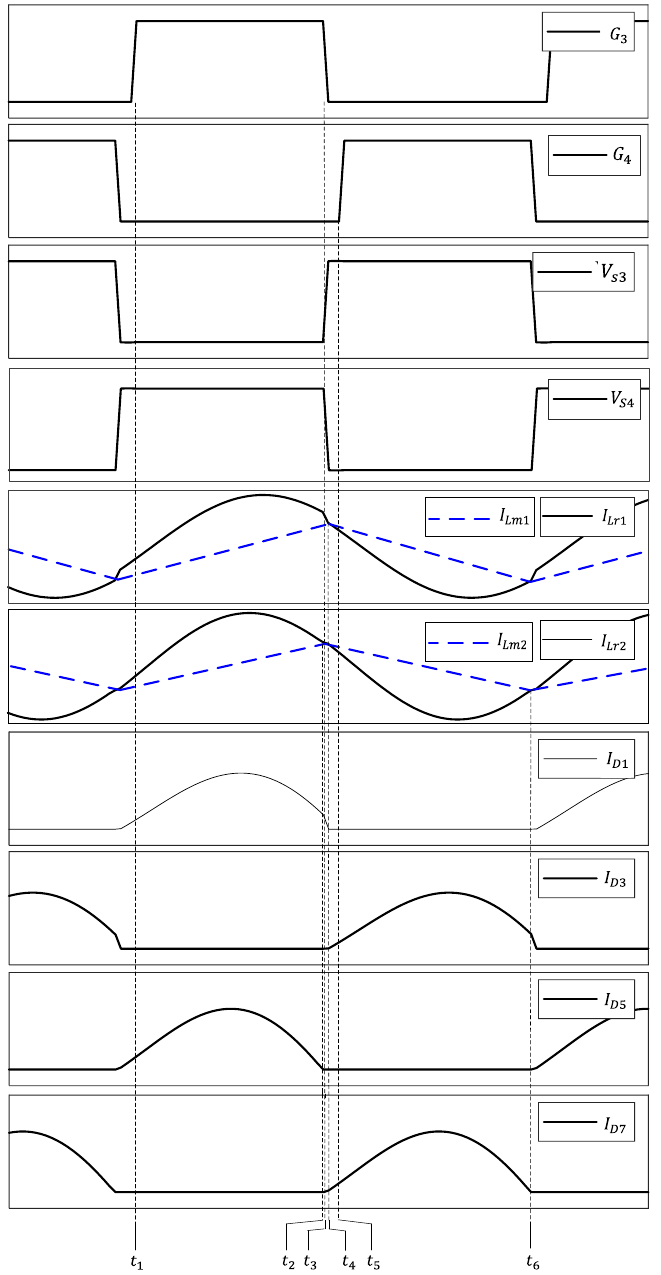}}

\vspace*{-0.4 cm}
\caption{Voltages and currents of the switches and diodes in steady-state operation.}

\vspace*{-0.6 cm}
\label{fig:waveforms}
\end{figure}

Fig.~\ref{fig:waveforms}. shows the current and voltage waveforms of the diodes, switches, and the magnetizing current during the operation modes. $G_1$, $G_4$, $V_{s3}$, and $V_{s4}$ are the gate drive commands and the switch voltages for $S_1$ and $S_4$ from the top, respectively. In this figure, $I_{Lm1}$, $I_{Lm2}$, $I_{Lr1}$, and $I_{Lr2}$ are the magnetizing current and inductor current for the transformers, respectively. During the charging interval, $I_{D1}$, $I_{D3}$, $I_{D5}$, and $I_{D7}$ show the output rectifier currents, respectively.

%\begin{equation}
%I_{lm}(t) =  \int_{t_1}^{t} V_{lm} \,dt =	 \frac{1}{l_m}V_{lm}(t-t_1)
%\end{equation}

% The resonance frequency of the tank resonance can be calculated from $f_r = \frac{1}{2\pi \sqrt{L_rC_r}}$

%\subsection{Operation Mode II: $t_2 \leq t<t_3$}
%In this operation mode, $S_1$ and $S_3$ are ON and $S_2$ and $S_4$ are OFF. In this mode, $I_\text{lr2}$ reaches to $I_\text{lm2}$ at $t_2$ and the secondary side of the $T_2$ gets zero. In this case, the magnetizing inductor of the primary side of this transformer forms a resonance with the tank resonance, which is calculated from $f_{sr2} = \frac{1}{2\pi \sqrt{(l_r+l_m)c_r}}$.

%\subsection{Operation Mode III: $t_3 \leq t<t_4$}
%In this mode, all switches are OFF as shown in Fig. . 
%In this mode, $I_\text{lr1}$ is greater than $Ilm1$ and the $C_\text{oss}$ are charged or discharged to achieve ZVS. The resonance tank current passes through the output capacitor of the switches, and based on the current direction, it charges the capacitor for $S_2$ and $s_3$ and discharges $s_1$ and $s_4$. 

%\subsection{Operation Mode IV: $t_4 \leq t<t_5$}
%In this mode, the body diode of $s_1$ and $s_4$ are ON and the output diodes $D_1$ and $D_2$ conduct. The direction of the inductor currents remain unchanged. During this interval, $I_lr1$ and $I_lr2$ pass through the body diode of $s_1$ and $s_4$. The $V_{DS}$ is, therefore, zero for $s_1$ and $s_4$. 

%\subsection{Switching Frequency Design}
%The LLC converter operates at a fixed frequency and close to the resonant frequency. The choice of switching frequency depends on the elements of the resonance tanks. Ideally, the resonant frequency of the two resonant tanks should be equal. However, the leakage inductance of the transformers are different in practice and the resonance tank components values may be different. This may be due to manufacturing tolerance or temperature impact. Based on the switching frequency, the resonant gain changes. This gain is also affected by the load. The resonant gain curve based on these two parameters is depicted in Fig. As load decreases, the gain increases. Moving toward the left side of the peak in these curves renders the converter to the capacitive region, which may damage the switches. To avoid damaging the switches, we must be on the right side of the peak or inductance region. The switching frequency in the design procedure is chosen to be smaller and close to the resonant frequency, which has inductive behavior. Therefore, the impact of load changes on resonant gain is reduced. 

\subsection{Resonant Tank Design}
This section discusses the designs of $L_r$ and $C_r$. The transformer leakage inductor $L_r$ affects the resonant tank gain and the resonance frequency. To reduce the effect of leakage inductance on the resonance frequency, $L_r$ should be large enough. The capacitor value is designed based on the resonance frequency and the maximum voltage stress across the capacitor in the resonant tank, which is calculated from:

\begin{equation} 
V_\text{cr}^\text{max}=I_\text{pri}^\text{peak}\sqrt{\frac{L_r}{C_r}}
\label{loadvoltage}
\end{equation}

\noindent where $I_\text{pri}^\text{peak}$ is the primary peak current. The resonance frequency is calculated using $F_\text{{r1,2}}=1/(2\pi \sqrt{L_rL_c})$. In practice, the resonant component values may not be equal for both transformers.

%\subsection{Control of the Proposed Topology}
%The primary side of the proposed topology works in a fixed duty cycle. Therefore, the output voltage of the converter is controlled by the secondary.

%=======================================================
\section{Simulation Results}
Simulation studies are performed in Plexim software to see the switches waveform and the zero-voltage switching operation under different rating power.
% Different scenarios are considered to validate the performance of the proposed topology.
%\subsection{Case I: Fast and Slow Chargers Are Active}
In the simulation study, both slow and fast chargers are working simultaneously. The simulation tests are done according to Fig.~\ref{fig:hybrid}, which shows one input and multiple outputs. For simplicity, only one of the parallel converters is included in the simulation studies.

%in which six 16~kW port are in parallel to deliver 96~kW. The slow charge power is 10~kW. The connected load for the slow charge port demands for 40~A with 400~V and for the second port, which is fast charge, the output current and voltage are 50~A and 200~V. The results for this scenario are shown in Fig.~\ref{fig:simulation5}. 
% As shown in this figure, the output current gets negative while the voltage drops, and then the current gets positive while the voltage is zero, indicating zero-voltage switching. These figures show the results for the output voltage and one of the switches voltage and current with power rating of 9~kW, 18~kW, and 26~kW from the top. 

\begin{figure}[t]
\centerline{\includegraphics[width= 0.98 \columnwidth ]{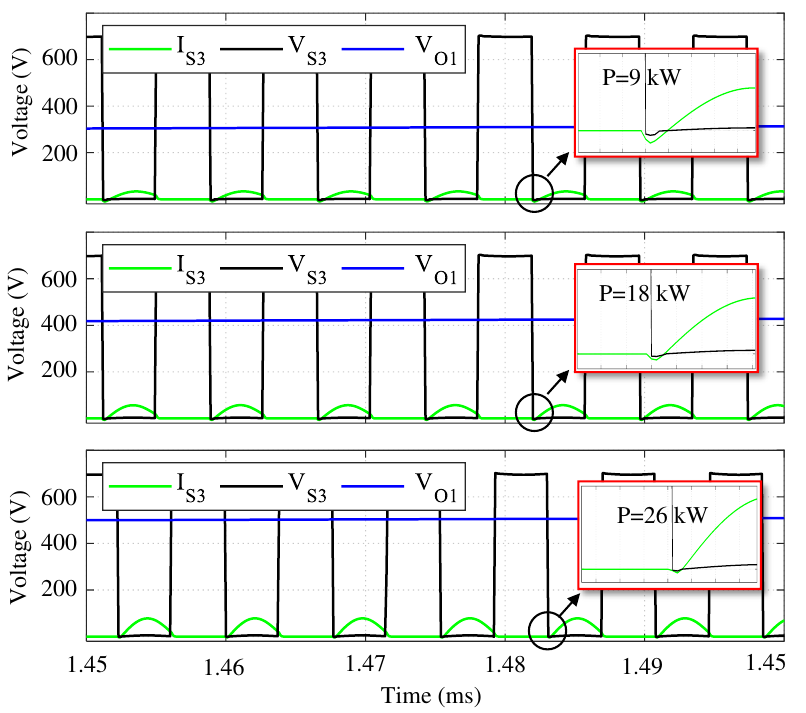}}

\vspace*{-0.3 cm}
\caption{Voltages and current of a switch at different rating power.}
\label{fig:int}
\end{figure}

% \subsection{Case I: Various Rating Power With One Port}
In this simulation study, one of the ports is examined under different rating powers to see the currents and voltages under ZVS. This study assumes that the first port provides power to the load and the second port is deactivated. In these cases, the converter delivers 9~kW, 18~kW, and 26~kW to validate under different ratings. The voltages and currents under these three rating powers are 303~V, 30~A, 425~V, 42.5~A, 510~V, and 51~A, respectively. The simulation results for these case studies are shown in Fig.~\ref{fig:int}. In this figure, $I_{s3}$ and $V_{s3}$ are the current and voltage for one switch of the primary side, and $V_{o1}$ is the output voltage of the rectified voltage on the secondary side. As shown in this figure, in all cases the proposed topology delivers power under zero-voltage switching.

\begin{table}[t]
\caption{System Parameters.}

\vspace*{-0.4 cm}
\begin{center}
\begin{tabular}{lcclcc}

\toprule
\textbf{Parameter} & \textbf{ } & \textbf{Value}  & \textbf{Parameter}  & \textbf{ } & \textbf{Value} \\ 
\midrule

\begin{tabular}[c]{@{}l@{}}Input \\ voltage\end{tabular}         & $V_{in}$              & \begin{tabular}[c]{@{}l@{}}150 V\\ \end{tabular} & \begin{tabular}[c]{@{}l@{}}Output\\ voltage \end{tabular}    & $V_O$              & 30-137~V         \\ 
 \midrule
\begin{tabular}[c]{@{}l@{}}Magnetizing \\ inductance \end{tabular} & $L_{m1,2}$              & 240~uH                                                 & \begin{tabular}[c]{@{}l@{}} Resonant  \\ Inductor\end{tabular}        & $L_{r1,2}$              &  31~uH         \\ 
\midrule
\begin{tabular}[c]{@{}l@{}}Resonant\\capacitor  \end{tabular}   & $C_{r1,2}$             & 60 nF                                                & \begin{tabular}[c]{@{}l@{}}Resonant\\ frequency\end{tabular}   & $F_r$              & 116 kHz        \\ 

 \bottomrule
\end{tabular}
\label{table:system_parameters}
\end{center}

\vspace*{-0.8 cm}
\end{table}

\section{Experimental Results}
\begin{figure}[t]

\vspace*{-0.2 cm}
\centerline{\includegraphics[width= 0.92\columnwidth ]{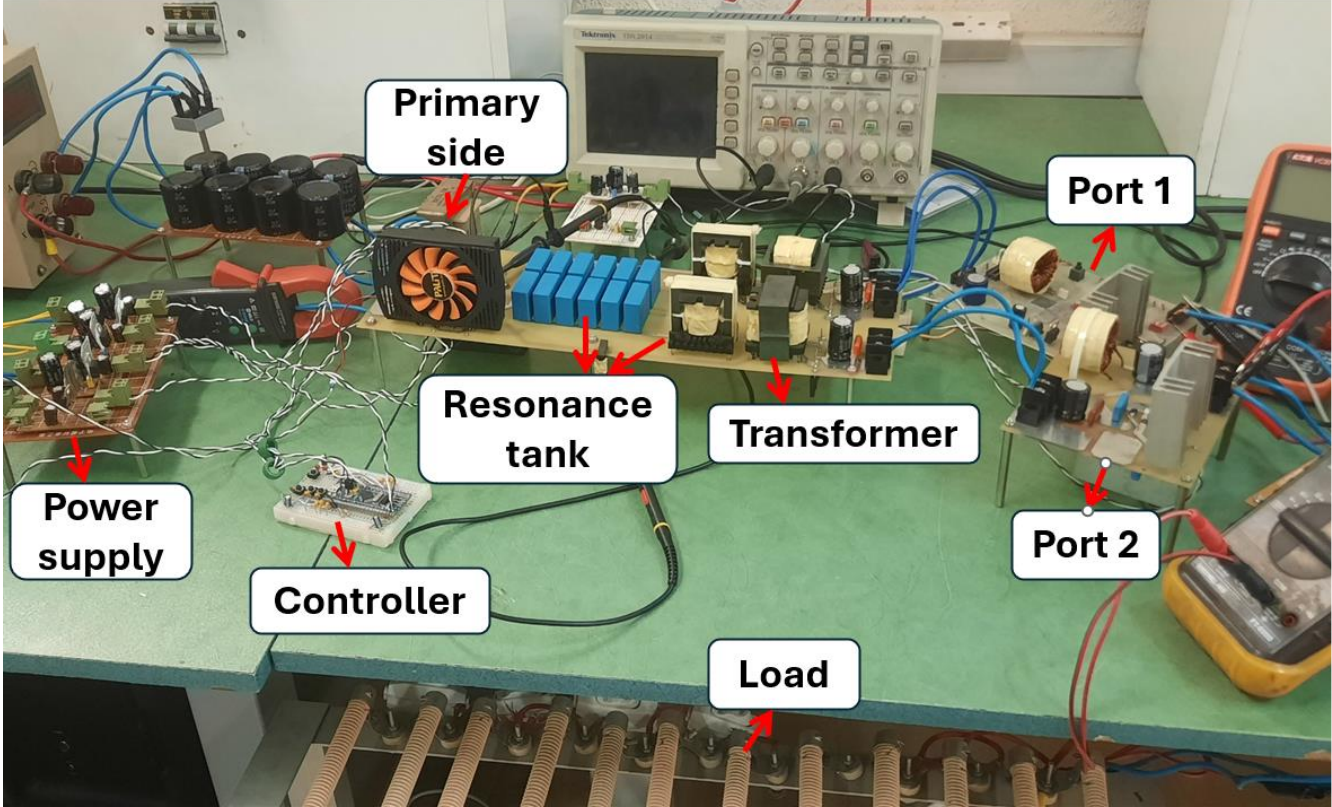}}

\vspace*{-0.35 cm}
\caption{Experimental setup.}
\label{fig:setup}
\end{figure}

A prototype is built in the lab to evaluate the proposed converter practically. The specification of the converter is shown in Table II. Fig.~\ref{fig:setup} shows the experimental setup. Two case studies are provided to show the converter efficiency and waveforms, as below. 

\subsection{Case I: $P_\text{out1}=460$~W and $P_\text{out2}=80$~W  }
In this case, the converter is examined to see the waveform and validate ZVS. The output power of the first port is assumed to deliver 460~W power to the load, and the second port provides 80~W to the load. In this test, the gate-source and gate driver voltages are captured. The result is shown in Fig.~\ref{fig:results}(a). As can be seen, the switches achieve zero-voltage switching. In this case study, the converter achieves 96\% efficiency.  

\subsection{Case II: $P_\text{out1}=80$~W and $P_\text{out2}=450$~W }
\begin{figure}[t]
\centerline{\includegraphics[width= 0.98\columnwidth ]{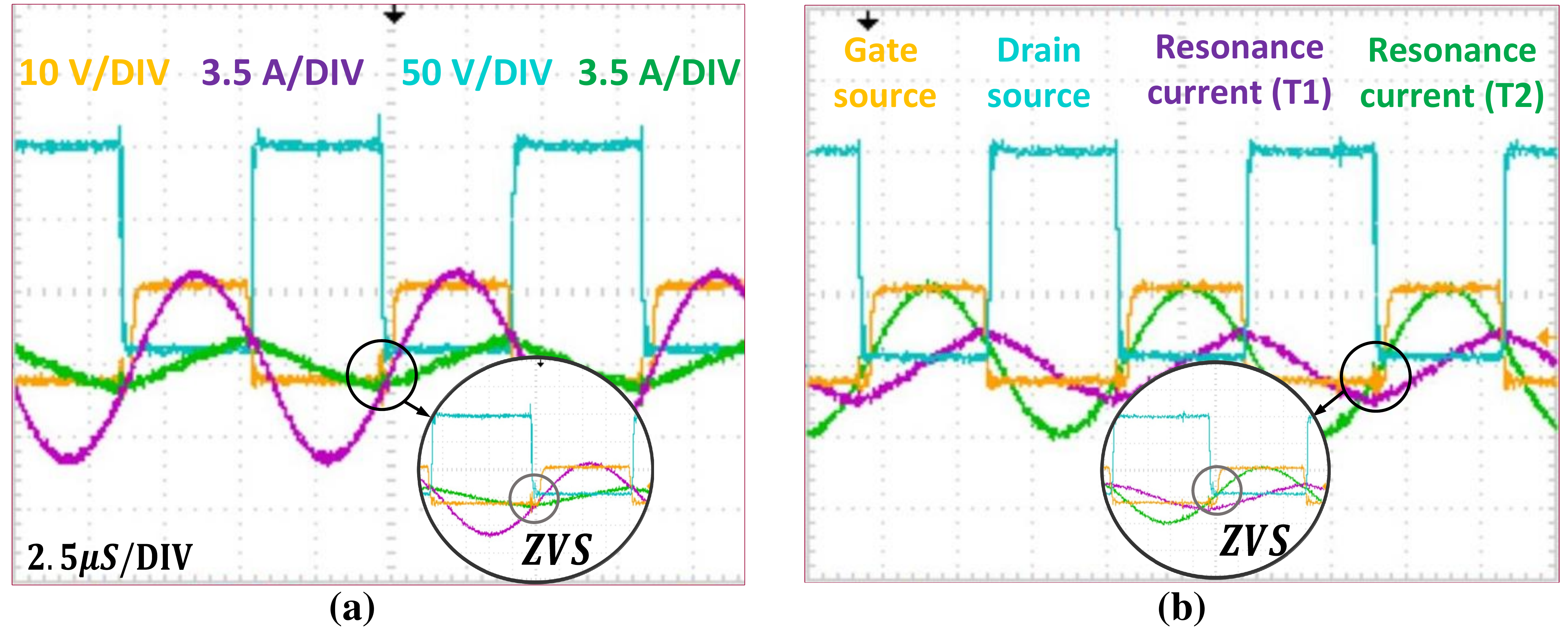}}

\vspace*{-0.34 cm}
\caption{Experimental results: (a) results for $P_\text{out1}=460$~W and $P_\text{out2}=80$~W, (b) results for $P_\text{out1}=80$~W and $P_\text{out2}=450$~W.}

\vspace*{-0.43 cm}
\label{fig:results}
\end{figure}

In this case, the converter is examined at another rating power. The output power of the first port is assumed to provide 80~W power to the load, and the second port delivers 450~W to the load in the case. In this test, the gate-source and gate driver voltages are captured. The result is shown in Fig.~\ref{fig:results}(b). As can be seen, the switches achieve zero-voltage switching. In this case study, the converter achieves 96\% efficiency using SiC-based switches.

\vspace*{-0.4 cm}

\section{Conclusion}
This paper proposes a hybrid three-port isolated DC-DC converter, which is based on the LLC resonant topology. This topology is designed specifically for hybrid EV charging stations. The proposed converter features both fast and slow charging ports. The topology uses an LLC resonant tank, providing isolation and enabling soft switching. By eliminating the need for the primary-side feedback signals, the complexity of the control circuit is significantly reduced. The proposed converter offers a lower number of components, which reduces the cost. In addition, it offers a simple control under fixed-frequency operation and ZVS. The efficacy of the converter is demonstrated through a 600 W prototype, validating the operation and its potential benefits for urban EV charging infrastructure. The converter achieves 96\% efficiency.

\section*{Acknowledgment}
\vspace*{-0.1 cm}

The authors would like to thank Kv Material, s.r.o company, for the financial support at the conference.

\bibliographystyle{IEEEtran}
\bibliography{ref.bib}

%==================================================
% \printbibliography %Prints bibliography

\end{document}